\documentclass[preprint,3p,authoryear]{elsarticle}
\usepackage{amssymb,amsmath}
\usepackage{graphicx}
\usepackage{dcolumn}
\usepackage{subfigure}
\usepackage{color}
\usepackage{soul}
\usepackage{float}
\usepackage{times}
\usepackage{soul}
\usepackage{xspace}
\usepackage[pdftex,breaklinks,colorlinks,
linkcolor=blue,
anchorcolor=blue,
citecolor=blue,
urlcolor=blue,]{hyperref}
\usepackage{lscape}
\usepackage[dvipsnames]{xcolor}
\usepackage{nameref}

\renewcommand{\eqref}[1]{\textup{{Eq.~(\ref{#1}})}}

\newcommand{\figref}[1]{\textup{{Figure~\ref{#1}}}}
\newcommand{\secref}[1]{\textup{{Section~\ref{#1}}}}
\makeatother

\makeatletter
\def\ps@pprintTitle{%
   \let\@oddhead\@empty
   \let\@evenhead\@empty
   \def\@oddfoot{\reset@font\hfil\thepage\hfil}
   \let\@evenfoot\@oddfoot
}
\makeatother

\begin{document}
\title{Universal Sampling Denoising (USD) for noise mapping and noise removal of non-Cartesian MRI}

\author[1,2]{Hong-Hsi Lee\corref{cor}}
\ead{HLEE84@mgh.harvard.edu}
\author[3]{Mahesh Bharath Keerthivasan}
\author[1,4]{Gregory Lemberskiy}
\author[1]{Jiangyang Zhang}
\author[1]{Els Fieremans}
\author[1]{Dmitry~S~Novikov}

\cortext[cor]{Corresponding author}

\address[1]{Center for Biomedical Imaging and Center for Advanced Imaging Innovation and Research (CAI2R), Department of Radiology, New York University School of Medicine, New York, NY 10016, United States}
\address[2]{Athinoula A. Martinos Center for Biomedical Imaging, Department of Radiology, Massachusetts General Hospital, Boston, Massachusetts, 02129, United States}
\address[3]{Siemens Medical Solutions USA, New York, NY 10016, United States}
\address[4]{Microstructure Imaging, INC, 600 Third Avenue, 2nd Floor, New York, NY 10016, United States}


\begin{abstract}
Random matrix theory (RMT) combined with principal component analysis has resulted in a widely used MPPCA noise mapping and denoising algorithm, that utilizes the redundancy in multiple acquisitions and in local image patches. RMT-based denoising relies on the uncorrelated identically distributed noise. This assumption breaks down after regridding of non-Cartesian sampling. 
Here we propose a Universal Sampling Denoising (USD) pipeline to homogenize the noise level and decorrelate the noise in non-Cartesian sampled k-space data after resampling to a Cartesian grid. 
In this way, the RMT approaches
become applicable to MRI of any non-Cartesian k-space sampling.  
We demonstrate the denoising pipeline on MRI data acquired using radial trajectories, including diffusion MRI of a numerical phantom and ex vivo mouse brains, as well as in vivo $T_2$ MRI of a healthy subject. The proposed pipeline robustly estimates noise level, performs noise removal, and corrects bias in parametric maps, such as diffusivity and kurtosis metrics, and $T_2$ relaxation time.
USD stabilizes the variance, decorrelates the noise, and thereby enables the application of RMT-based denoising approaches to MR images reconstructed from any non-Cartesian data. In addition to MRI, USD may also apply to other medical imaging techniques involving non-Cartesian acquisition, such as PET, CT, and SPECT.

\end{abstract}
\date{\today}
\maketitle

\section{Introduction}
The development of MRI enables in vivo evaluation of tissue properties through various signal contrasts, such as diffusion MRI (dMRI) and relaxation time mapping, in which inference is made based on measured signal attenuation in multiple images \citep{jones2010book,novikov2019review}. 
The signal-to-noise ratio (SNR), which decreases with stronger signal attenuation such as in dMRI, can become quite low;  thermal noise corrupts image quality and propagates into estimated parametric maps. For example, noise in diffusion tensor imaging (DTI) leads to eigenvalue repulsion in the diffusion tensor \citep{pierpaoli1996repulsion}, resulting in over-estimated axial diffusivity and fractional anisotropy, and an under-estimated radial diffusivity.

In this work, we consider a family of algorithms  proposed for noise reduction assuming spatially and temporally independent and identically distributed (i.i.d.) white Gaussian noise. 
Such algorithms were originally developed for Cartesian sampled data.  
Using the redundancy in MRI data of multiple contrasts via a local or non-local image patch, it is possible to identify \citep{veraart2016noisemap} and remove \citep{veraart2016denoising} the noise components in the domain of principal component analysis (PCA), where the pure noise-related eigenvalues obey the universal Marchenko-Pastur (MP) law \citep{marchenko1967MP} based on the random matrix theory (RMT) results for large noisy covariance matrices.  
The corresponding MPPCA denoising algorithm was first applied to magnitude dMRI data. \citep{veraart2016noisemap,veraart2016denoising} The value of MPPCA was shown in other redundant acquisitions such as $T_2$ mapping\citep{does2019mppca} and fMRI\citep{ades2021fmri,vizioli2021nordicfmri}, and it became the first step in image processing pipelines.\citep{ades2018designer} 

The complex i.i.d. noise assumption of a fully-sampled MR image breaks down in parallel imaging (due to a non-unitary transformation of the i.i.d. Gaussian noise in receive coils \citep{pruessmann1999sense,pruessmann2001correlation,robson2008pisnr,breuer2009gfactor})
or after taking the absolute value\citep{does2019mppca}. 
Hence, to further improve the performance of RMT-based denoising, the effect of spatially varying noise level (geometry factor, g-factor) \citep{moeller2021nordic,vizioli2021nordicfmri} or full noise correlation between coils \citep{aja2014paraimag,cordero2019denoise} due to linear transformations in parallel imaging  should be considered in reconstructed images. 
One way to preserve the i.i.d. noise statistics is to denoise all aliased coil images (after applying Fourier transform to the acquired k-space lines) before image reconstruction and coil combination. 
Denoising such images before image reconstruction prevents the problem of spatial varying noise level and noise correlation between coils, effectively leading to a 5-fold decrease in the Rician noise floor \citep{lemberskiy2019rmt,lemberskiy2021vcc}.

\begin{figure}[t!]
\centering
	\includegraphics[width=0.75\textwidth]{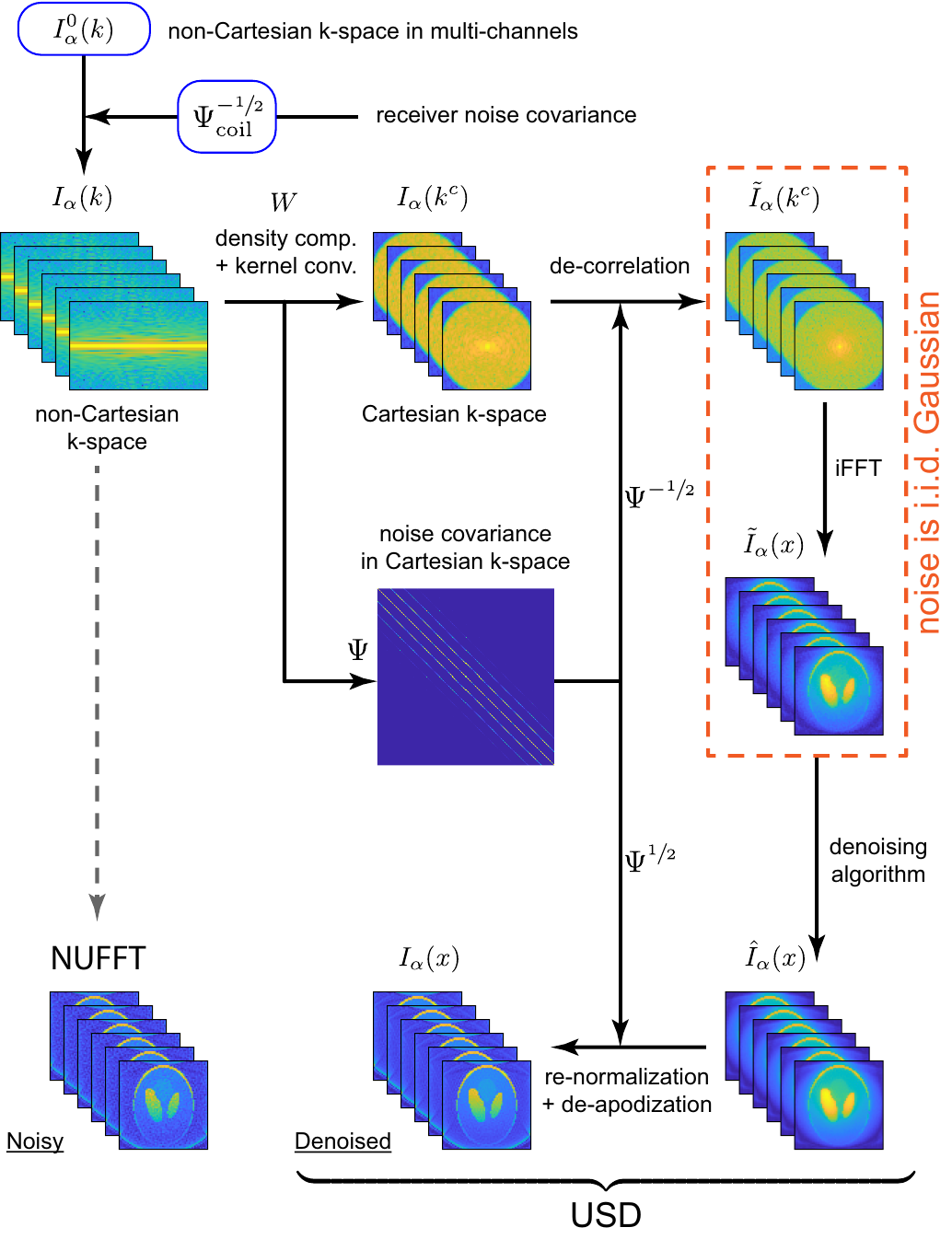}
	\caption{The pipeline of Universal Sampling Denoising (USD) for non-Cartesian k-space data, as detailed in the Theory section, illustrated here for a radial trajectory. After removing any noise correlation between receiver channels by using noise scans (\secref{sec:theory-receiver}), the key idea of USD is to further de-correlate the noise statistics in the gridded Cartesian k-space prior to applying any denoising algorithm in the image space (\secref{sec:theory-pipeline}). The noise covariance matrix $\Psi=WW^H$ in Cartesian k-space is determined by coefficients $W$ of density compensation and convolution with interpolation kernel, \eqref{WW}. After denoising using the MPPCA algorithm, re-normalization of the denoised image recovers the original image contrast.}
	\label{fig:usd-pipeline}
\end{figure}

Non-Cartesian sampling in k-space provides flexibility in MRI acquisitions, e.g., for echo time (TE) shortening or motion robustness \citep{block2014stackstars,wilm2020spiral}, and it has been used to achieve highly accelerated image acquisition, such as Golden-angle radial sparse parallel (GRASP) MRI \citep{feng2014grasp}. Although noise level estimation and noise removal are equally essential for image reconstruction and quantitative analysis based on non-Cartesian data, denoising non-Cartesian sampled MRI is challenging. 
Indeed, while different k-space trajectories (such as radial spokes) have i.i.d. noise in k-space, ``patching" them (or their images) together may not provide sufficient redundancy. Ideally, one would like to utilize spatial redundancy and denoise in the (reconstructed) image space, but this requires the additional interpolation of non-uniformly sampled Fourier data onto a Cartesian grid in the reconstruction. This non-unitary linear transformation introduces a spatially varying noise level and noise correlations between each voxel or a k-space data point (\figref{fig:usd-pipeline}). This challenge is even greater than that arising in denoising Cartesian undersampled data. Failure to address these noise correlations results in image blurring after denoising (\figref{fig:sim-dwi-md}). 

Here, to estimate and remove noise for any sampling scheme, we propose the universal sampling denoising (USD) pipeline to simultaneously homogenize the noise level (i.e., variance stabilization) and de-correlate the noise in the gridded k-space and image space for any non-Cartesian sampling, such that MPPCA can be applied to identify and remove the noise in the Cartesian image domain after the regridding. We demonstrate the pipeline in radially sampled diffusion MRI data from numerical phantoms and ex vivo mouse brains, and radially sampled in vivo $T_2$ MRI data of human abdomen.

\section{Theory} \label{sec:theory}
We propose the Universal Sampling Denoising (USD) pipeline to denoise the non-Cartesian data, summarized in \figref{fig:usd-pipeline}. USD de-correlates the noise before applying MPPCA denoising, along the following steps. 

First, noise correlations in receiver channels are removed by using the noise covariance matrix given by the noise prescan acquired without radiofrequency (RF) pulses \citep{roemer1990array,kellman2005snr}. 
Next, the remaining noise correlations between voxels due to non-uniform resampling onto a Cartesian grid are removed by using the noise covariance matrix determined by coefficients of density compensation and kernel convolution in non-uniform fast Fourier transform (NUFFT). This is the nontrivial USD step. After noise de-correlation, the noise is approximately i.i.d. Gaussian and is subsequently removed by applying MPPCA denoising. Finally, the denoised images are re-normalized to restore the image contrast. 

\subsection{Noise de-correlation in receiver channels} \label{sec:theory-receiver}

Consider the noisy acquisition (image plus complex-valued noise) $I^0_\alpha(k) + \epsilon_{\alpha}(k)$
for the RF receive channel $\alpha$, where $k$ is the k-space point in any (in general, non-Cartesian) trajectory. 
The ``original'' noise $\epsilon_{\alpha}$  
is correlated along the RF coil dimension $\alpha$, such that the expectation value
\begin{equation} \label{I-orig}
    \langle \epsilon_{\alpha}(k) \epsilon^*_{\beta}(k')\rangle =
     \delta_{kk'}\cdot {\Psi_\text{coil,}}_{\alpha\beta} \,, 
\end{equation}
and yet is independent of $k$ (hence the Kronecker $\delta_{kk'}$). 
The coil covariance matrix $\Psi_\text{coil}$, that can be determined using noise prescan without RF excitations, defines noise correlations in receiver channels \citep{roemer1990array,kellman2005snr}. 

As the first step of the USD pipeline, the measured signal 
is de-correlated via \citep{roemer1990array,kellman2005snr}
\begin{equation} \label{decorr-coil}
I_\alpha(k) + \varepsilon_\alpha(k) = 
\sum_\beta \left(\Psi_\text{coil}^{-1/2}\right)_{\alpha\beta} 
\cdot \left(I^0_\beta(k) + \epsilon_{\beta}(k)\right), 
\end{equation}
such that the noise $\epsilon\to\varepsilon$ becomes i.i.d. complex-valued Gaussian, 
\begin{equation} \label{eq:decorr-coil-noise}
\langle \varepsilon_{\alpha}(k) \varepsilon^*_{\beta}(k')\rangle =   \delta_{kk'}\cdot \delta_{\alpha\beta}\,, 
\end{equation}
having no  correlations between the ``rotated'' coils. 





\subsection{Universal denoising pipeline for any sampling scheme} \label{sec:theory-pipeline}


In this work, we consider the case when 
the original image  is acquired at the non-Cartesian $k$-points. Since an MR image is defined (in the $x$-space) on a Cartesian grid, the non-Cartesian k-space points have to be first regridded (interpolated) onto the Cartesian k-space grid. The combination of regridding in k-space and the inverse fast Fourier transform (iFFT) into the $x$-space is referred to as the non-uniform FFT (NUFFT). Here we use $1/\sqrt{N}$ as the normalization factor of FFT and iFFT to ensure that they are unitary. Since FFT is a unitary operation, all the noise correlations introduced by the NUFFT come from the k-space regridding. 

USD is meant to remove such correlations. For that, consider the four conventional NUFFT steps \citep{jackson1991nufft}:
\begin{enumerate}
    \item Density compensation in k-space;
	\item Convolution with an interpolation kernel $C(k)$ in k-space, e.g., the Kaiser-Bessel function;
	\item iFFT to image space; and
	\item De-apodization in image space.
\end{enumerate}
The first two steps of NUFFT can be considered as a linear non-unitary transformation (same for each coil $\alpha$):
\begin{equation} \label{eq:Ik-C-NC}
    I_\alpha(k^c_i) + \varepsilon_\alpha(k^c_i) =\sum_j w_{ij}\cdot \left[I_\alpha(k_{j}) + \varepsilon_\alpha(k_{j})\right],
\end{equation}
where $k_i^c$ refer to the Cartesian k-space points, as opposed to general non-Cartesian $k_j$ acquired originally. 
The weights $W=(w_{ij})=(d_j\cdot C(k_j-k_i^c))$ incorporate density compensation $d_j$ and kernel convolution $C(k_j-k_i^c)$. 
Typically, $w_{ij}$ is a local kernel, involving a few adjacent non-Cartesian points to interpolate a Cartesian one. 
Note that indices $i$ and $j$ label the points with the 2- or 3-dimensional coordinates. Hence, they run up to the total number of k-space points. 
For the most general form of \eqref{eq:Ik-C-NC} and the following denoising pipeline, the linear transformation weights $w_{ij}$ can be extended to not only the dimension of k-space or image space, but also the dimension of coils, slices, or temporal domain for parallel imaging, simultaneous multi-slice, or any other fast imaging techniques.
This is out of the scope of this study, and we reserve these directions for the future.

The regridding (\ref{eq:Ik-C-NC}) introduces noise correlations on the Cartesian grid, similar to Refs.~ \citep{sengupta1999noisecorr,pruessmann1999sense,robson2008pisnr,breuer2009gfactor}:
\begin{equation} \label{WW}
    \langle \varepsilon_\alpha(k^c_i) \varepsilon_\beta^*(k^c_j) \rangle = \sigma^2_0\cdot \delta_{\alpha\beta} \cdot \Psi_{ij}\,,
    \quad
    \Psi= WW^H\,,
\end{equation}
with $\sigma_0$ the noise level in the non-Cartesian k-space. Based on \eqref{eq:decorr-coil-noise}, $\sigma_0=1$ when the coil data is de-correlated along the RF coil dimension using noise prescan. From now on, we assume that the noise level is homogeneous and constant (not necessarily unity) in the acquired non-Cartesian k-space. 
The key observation is that the noise covariance matrix $\Psi$ on the k-space Cartesian grid is determined by the weighting coefficients $W$.

With the knowledge of how noise gets correlated, the k-space data on the Cartesian grid can be de-correlated, analogously to \eqref{decorr-coil}:
\begin{equation} \label{decorr-k}
    \tilde{I}_\alpha(k^c_{i}) + \tilde{\varepsilon}_\alpha(k^c_{i}) = \sum_j \left(\Psi^{-1/2}\right)_{ij} \cdot \left[I_\alpha(k^c_j) + \varepsilon_\alpha(k^c_j)\right],
\end{equation}
where $\tilde{I}$ and $\tilde{\varepsilon}$ are the k-space signal and noise on the Cartesian grid after de-correlation.

While the de-correlation (\ref{decorr-k}) changes the local contrast in the k-space, it ensures the i.i.d. complex Gaussian noise, 
\begin{equation} \label{eq:epsilon}
    \langle \tilde{\varepsilon}_\alpha(k^c_{i})\tilde{\varepsilon}^*_\beta(k^c_j)\rangle =
    \sigma_0^2\cdot \delta_{\alpha\beta}\cdot \delta_{ij} \,.
\end{equation}
Likewise, the corresponding signal contrast in the noise de-correlated image $\tilde{I}_\alpha(x)$ is very different from that of the ground truth coil image, and yet, after the unitary iFFT transformation, the corresponding image noise $\tilde\varepsilon_\alpha(x)$ remains spatially-uncorrelated i.i.d. Gaussian. 
Hence, the image noise statistics in the ``tilde'' space make it particularly suitable for applying RMT-based noise removal algorithms such as MPPCA. 

After de-correlation and iFFT, we are left with the three-way MRI data matrix sampled in a patch $\Omega(x_0)$ around the voxel $x_0$: ${\cal X}_{\alpha x m}(\Omega) = \left[\tilde I_{\alpha,m}(x) + \tilde\varepsilon_{\alpha,m}(x)\right]|_{x\in\Omega(x_0)}$, where $m$ labels independent measurements (e.g., diffusion $q$-space points, or  images in a time-series). The noise $\tilde\varepsilon_{\alpha,m}(x)$ is uncorrelated and i.i.d. Gaussian across all three indices in ${\cal X}_{\alpha x m}$. One can then denoise such an object using RMT-based methods by ``flattening'' it along different dimensions. Empirically, combining the coil dimension with the patch around the voxel has the best performance \citep{lemberskiy2019rmt,lemberskiy2021vcc}, i.e., re-arranging ${\cal X}_{\alpha x m}$ as an $(N_\alpha\cdot N_x)$ by $N_m$ matrix, with $N_{(\cdot)}$ denoting the number of elements along the dimension $(\cdot)$. 

Alternatively, one can first combine the coil images, obtaining the  matrix ${\cal X}_{xm}$ of size $N_x$ by $N_m$, and then apply MPPCA complex denoising to this two-way object, in which case an additional noise variance stabilization would be required for the spatially varying coil combination weights.

RMT-based MPPCA denoising is then applied to estimate the noise level $\hat{\sigma}(x)$ and the number $P(x)$ of signal components in the PCA domain in the patch $\Omega(x)$ (either a local square patch, or a non-locally chosen, e.g., based on signal similarity) around voxel $x$, and to remove the noise \citep{veraart2016denoising,veraart2016noisemap,lemberskiy2019rmt,lemberskiy2021vcc}. 
The data of all coils are denoised jointly in the following steps: phase demodulation, virtual coil compression, and PCA thresholding based on MP distribution. The noise components are removed by using the optimal shrinkage of singular values of PCA \citep{gavish2017shrinkage}. After denoising, the noise variance decreases by a factor of
$P\cdot(1/N + 1/M)$, assuming $P\ll M, N$, such that SNR at voxel $x$ increases by a factor $\approx \sqrt{M/P(x)}$, where $M$ is the smaller dimension of data matrix defined at the voxel \citep{veraart2016noisemap}. In most cases, $M$ is the number of scanned images.  

Finally, the denoised image $\hat{I}_\alpha(x)$ is re-normalized to recover its original contrast:
\begin{equation} \label{eq:renorm}
    I_\alpha(x) = \frac{1}{c(x)} F^H\cdot\Psi^{1/2}\cdot F\cdot \hat{I}_\alpha(x)\,,
\end{equation}
where $c(x)$ is the FT of convolution kernel $C(k)$ for de-apodization, and $F=(F_{k_i^c x}) \equiv (F_{ix})$ denotes the fast Fourier transform. 
The denoised images $I_\alpha(x)$ in multiple channels are then adaptively combined \citep{walsh2000adaptcomb,griswold2002normalize},
\begin{equation} \label{eq:adapt-comb}
    I(x)=\sum_\alpha p_\alpha(x) \cdot I_\alpha(x)\,,
\end{equation}
with spatially varying adaptive combination weights $p^T=p_\alpha(x)$. 

The noise map $\hat{\sigma}(x)$ of de-correlated image
$\tilde{I}_\alpha(x)$ is an estimate of $\sigma_0$ in \eqref{eq:epsilon}, which can be translated into the noise map $\sigma(x)$ of images in the original contrast without denoising, via \citep{breuer2009gfactor}
\begin{equation} \label{eq:noise-usd}
\begin{split}
    \sigma(x) &= \frac{1}{c(x)}\sqrt{\left|F^H\cdot\Psi\cdot F\right|_{x,x}}\cdot \sigma_0 \\
    &= \frac{1}{c(x)}\sqrt{\left|\sum_{ij} F_{xi}^*\cdot\Psi_{ij}\cdot F_{jx}\right|} \cdot\sigma_0\,,
\end{split}
\end{equation}
with $|\cdot|_{x,x}$ denoting the absolute value of diagonal element at voxel $x$ (Appendix~\ref{sec:noise-level-transform}).
Given that the noise in de-correlated image $\tilde{I}_\alpha(x)$ is i.i.d. Gaussian, its noise variance $\sigma_0^2$ can be estimated by the $\hat{\sigma}^2(x)$ or its average over space $\langle\hat{\sigma}^2(x)\rangle$, with $\langle\cdot\rangle$ denoting averaging over space.
Similarly, substituting \eqref{eq:renorm} into \eqref{eq:adapt-comb} and following the derivation in Appendix~\ref{sec:noise-level-transform}, the noise level of combined image is given by
\begin{equation} \label{eq:noise-usd-comb}
\begin{split}
    \sigma_\text{comb}(x) &= \frac{1}{c(x)}\sqrt{\left| (F\cdot p^*)^H \cdot \Psi \cdot (F \cdot p^*) \right|_{x,x}} \cdot \sigma_0 \\
    &= \frac{1}{c(x)}\sqrt{\left| \sum_{\alpha ij} p_\alpha(x)\cdot F_{xi}^* \cdot \Psi_{ij} \cdot F_{jx} \cdot p^*_\alpha(x) \right|} \cdot\sigma_0\,.
\end{split}
\end{equation}

The first lines in Equations~(\ref{eq:noise-usd}) and (\ref{eq:noise-usd-comb}) suggest the most general form of the noise level transformation, where the noise correlation $\Psi$ can be extended to not only the dimension of k-space or image space, but also the dimension of coils, slices, or temporal domain for parallel imaging, simultaneous multi-slice, or any other fast imaging techniques.
In this study, we assume that  parallel imaging and simultaneous multi-slice are not applied, yielding the second lines in Equations~(\ref{eq:noise-usd}) and (\ref{eq:noise-usd-comb}); thus, the adaptive combination weight $p$ is separable from the FFT coefficient $F$ and noise covariance matrix $\Psi$ of NUFFT since the later two are independent of the coils, leading to
\begin{equation}
    \sigma_\text{comb}(x)=\frac{1}{c(x)}\sqrt{|p^T\cdot p^*|_{x,x}}\cdot\sqrt{\left| F^H \cdot \Psi \cdot F \right|_{x,x}} \cdot \sigma_0\,.
\end{equation}
Using \eqref{eq:noise-usd}, we can define the g-factor for the coil image due to NUFFT:
\begin{equation}
    g
    = \frac{\text{SNR}^\text{full}}{\text{SNR}^\text{nufft}}
    = \frac{\sigma(x)}{\sigma_0}
    = \frac{1}{c(x)}\sqrt{\left|F^H\cdot\Psi\cdot F\right|_{x,x}}\,.
\end{equation}
Similarly, the g-factor of the combined image is given by
\begin{equation}
    g_\text{comb}
    = \frac{\text{SNR}^\text{full}_\text{comb}}{ \text{SNR}^\text{nufft}_\text{comb}}
    = \frac{\sigma_\text{comb}(x)}{\sigma_0} \cdot \frac{1}{\sqrt{|p^T\cdot p^*|_{x,x}}}\,,
\end{equation}
and $g_\text{comb}=g$ when  parallel imaging is not applied.

\section{Methods}
We demonstrated the USD pipeline on radially sampled k-space dMRI data of a numerical phantom and ex vivo mouse brain, and in vivo $T_2$ MRI data of human abdomen.

\subsection{Numerical simulation}
To demonstrate USD in simulated data, we created a 2-dimensional Shepp-Logan phantom of size 64×64, including a non-diffusion-weighted-image (non-DWI) $S_0\in[0,1]$ and 30 DWIs of b-value $b = 0.1-1$ ms/$\mu$m\textsuperscript{2} with 12 different coils of linear coil sensitivity. The multiple channels provided extra data redundancy for noise removal, and were adaptively combined after denoising. The DWI signal was $S=S_0\cdot \exp(-bD)$ with a diffusivity map $D=\left|6S_0-4S_0^2\right|$. Its k-space data was sampled on radial trajectories, consisting of 100 spokes/image and 64 sampled data points/spoke. Gaussian noise with standard deviation $\sigma(k_{NC})=0.05$ was added in the real and imaginary part of k-space data on radial trajectories. The non-Cartesian diffusion weighted images (DWIs) were reconstructed by using the NUFFT toolbox \citep{fessler2007nufft}, and the diffusivity map was estimated by fitting a mono-exponential fit.

\subsection{Ex vivo mouse brain data}
Further, we demonstrated USD in dMRI data of an ex vivo mouse brain controlled at 36°C during the scan. dMRI measurements were performed using a monopolar pulsed-gradient 2-dimensional center-out radial acquisition \citep{pauly1989T2} on a 7 Tesla MRI system (Bruker Biospin, Billerica, MA, USA) with a 4-channel receive-only cryocoil in combination with a 72-mm diameter volume coil for excitation, providing extra data redundancy for noise removal, and adaptively combined after denoising. We obtained 4 non-DWIs and 60 DWIs of b-value $b= [1, 2]$ ms/$\mu$m\textsuperscript{2} along 30 directions per b-shell, with a voxel size = 0.156×0.156×1 mm\textsuperscript{3}, FOV = 20×20 mm\textsuperscript{2}, and fixed TE/TR = 20/400 ms. The ground truth image was reconstructed from 402 center-out radial spokes with 70 data points/spoke by using the NUFFT toolbox \citep{fessler2007nufft}, and averaged over two repeated measurements. In addition, the noisy raw data was reconstructed from only 201 spokes without averaging over repeats. Hence, the SNR of the ground truth is higher than that of the noisy raw data. The USD denoising pipeline was applied to noisy raw data. Voxelwise kurtosis tensor fitting \citep{jensen2005dki,veraart2013wlls} was performed to the ground truth, noisy raw data, and denoised data to extract parametric maps of diffusion and kurtosis tensor metrics, such as mean diffusivity (MD), axial diffusivity (AD), radial diffusivity (RD), colored fractional anisotropy (FA), mean kurtosis (MK), axial kurtosis (AK), and radial kurtosis (RK).

\subsection{In vivo human data}
Finally, we demonstrated USD in in vivo $T_2$ relaxometry data of a human abdomen. Following IRB approval for prospective data collection and informed consent, a volunteer (female, 39 years old) was imaged using a 6-channel surface array coil and an 18-channel spine coil on a commercial MRI system (MAGNETOM Aera 1.5T, Siemens Healthcare, Erlangen, Germany) modified to operate as a prototype scanner at a field strength of 0.55T. Data was acquired using a 2-dimensional radial turbo spin echo sequence  \citep{altbach2002radtse} with the following parameters: TE = 7.9-158 ms (echo train length = 20, echo spacing = 7.9 ms), voxel size = 1.48$\times$1.48$\times$6 mm\textsuperscript{3}, TR = 3.3 s, and refocusing flip angle = 180$^\circ$. Each image of matrix size 256$\times$256 was reconstructed from 51 radial spokes with 512 data points/spoke by using the NUFFT toolbox \citep{fessler2007nufft}. The USD denoising pipeline was applied, and the $T_2$ value in each voxel is decided by matching the denoised signal with a dictionary of multi-spin-echo $T_2$ relaxation, built based on the extended phase graph method \citep{weigel2015epg} with a fixed $T_1=1000$ ms, $T_2=1-200$ ms, and $B_1^+$ $= 60-120\%$ \citep{lebel2010transverse}.

\section{Results}
\subsection{Numerical simulation}
\begin{figure}[t!]
\centering
	\includegraphics[width=0.75\textwidth]{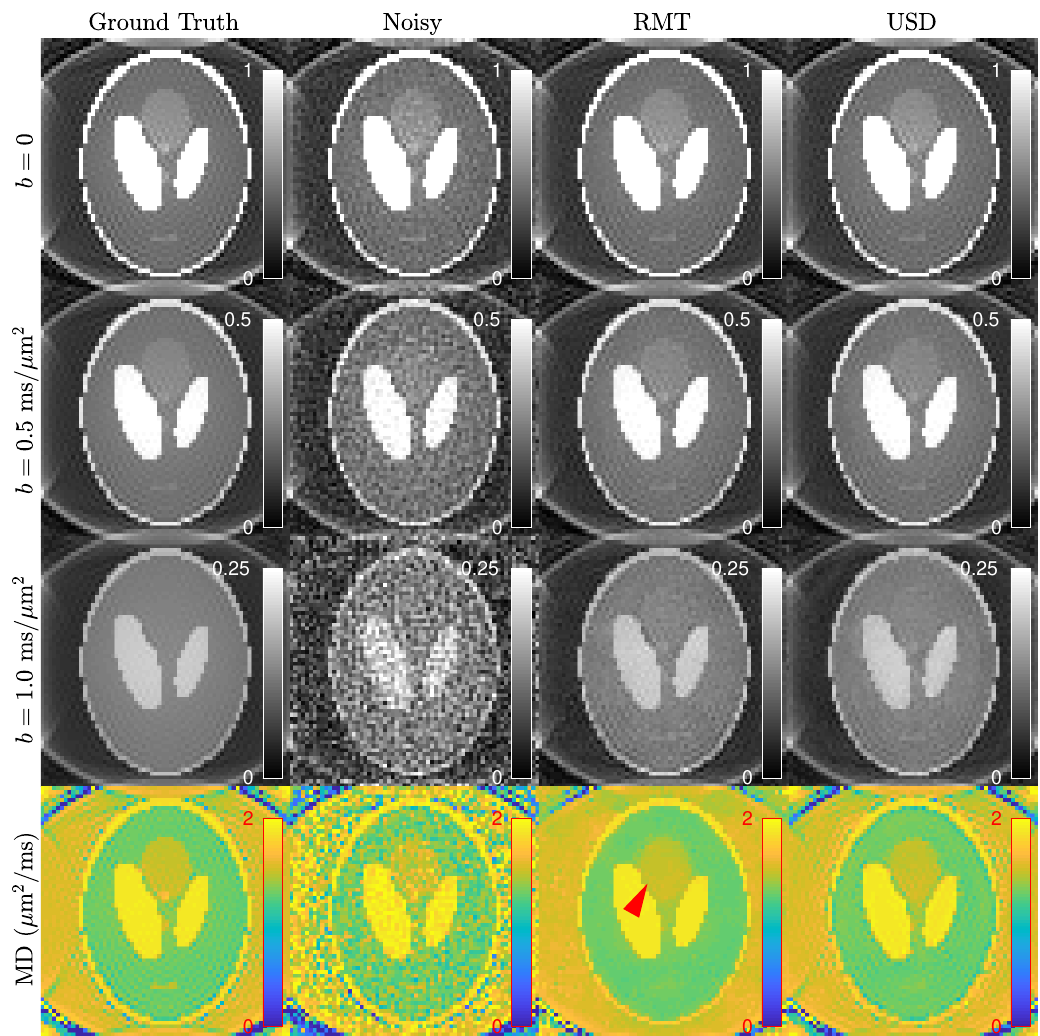}
	\caption{Numerical simulation of a diffusion phantom (1 $b=0$ image + 30 DWIs). The ground truth (GT) and noisy data were reconstructed by applying NUFFT to the original k-space data on radial trajectories. The noise in noisy data was significantly reduced by using either RMT-based denoising without noise de-correlation or the USD pipeline. The RMT without noise de-correlation seemed to reduce the noise in images, yet lead to blurring in the MD map (red arrow). In contrast, the USD successfully recovered the MD values without excessive smoothing.}
	\label{fig:sim-dwi-md}
\end{figure}

\begin{figure}[t!]
\centering
	\includegraphics[width=0.50\textwidth]{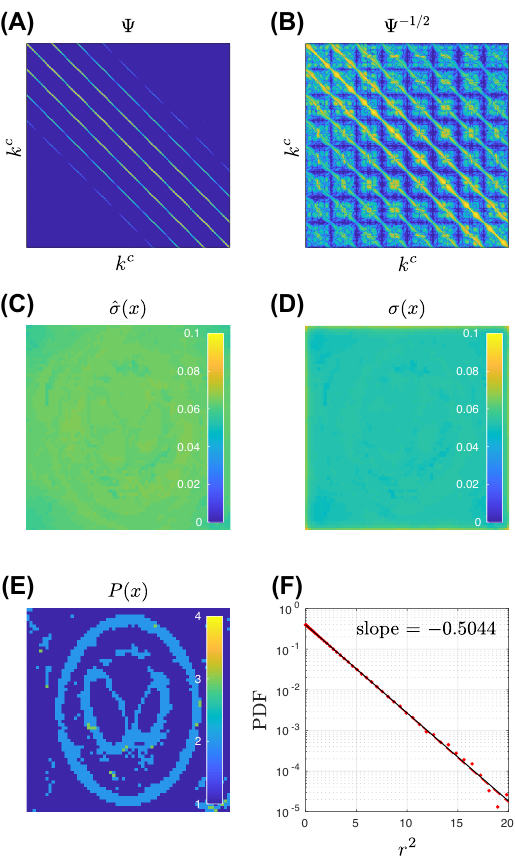}
	\caption{Noise covariance matrix in gridded k-space and noise statistics in numerical simulations (\figref{fig:sim-dwi-md}). (A-B) In the Cartesian k-space $k^c$, the noise covariance matrix $\Psi$ was sparse, but its square root of inverse $\Psi^{-1/2}$ was not. (C-D) The noise maps before and after re-normalization, $\hat{\sigma}(x)$ and $\sigma(x)$, were both smooth. (E) The number $P(x)$ of signal components in PCA domain was low, indicating the applicability of RMT-based denoising algorithm to noise de-correlated images. (F) The normalized image residual $r$ in \eqref{eq:residual} was almost normally distributed, manifested by the slope $\approx-1/2$ in the semi-log scale.}
	\label{fig:sim-noise}
\end{figure}

In the numerical phantom, USD substantially reduced the noise and recovered the true values from the diffusivity values biased due to the noise floor in noisy raw data (\figref{fig:sim-dwi-md}). Interestingly, the noise covariance matrix due to NUFFT was sparse, and yet its inverse square root was not (\figref{fig:sim-noise}A-B). The noise map before and after re-normalization was smooth (\figref{fig:sim-noise}C-D), and the number of signal components in PCA domain was low (\figref{fig:sim-noise}E).

For quantitative analysis, we define the normalized image residual in multiple channels,
\begin{equation} \label{eq:residual}
    r\equiv \left[ \hat{I}_\alpha(x)-\tilde{I}_\alpha(x)\right]/\hat{\sigma}(x)\,.
\end{equation}
When the denoising algorithm removes most of the noise without corrupting the signal, the $\hat{\sigma}$-normalized residual $r$ should be normally distributed. In simulations, the residual histogram in the semi-log scale is below the reference line of slope $-1/2$, i.e., PDF$\sim\exp\left(-\tfrac{1}{2}r^2\right)/\sqrt{2\pi}$, indicating the applicability of USD in simulations (\figref{fig:sim-noise}F).

\subsection{Ex vivo mouse brain data}
\begin{figure}[t!]
\centering
	\includegraphics[width=0.99\textwidth]{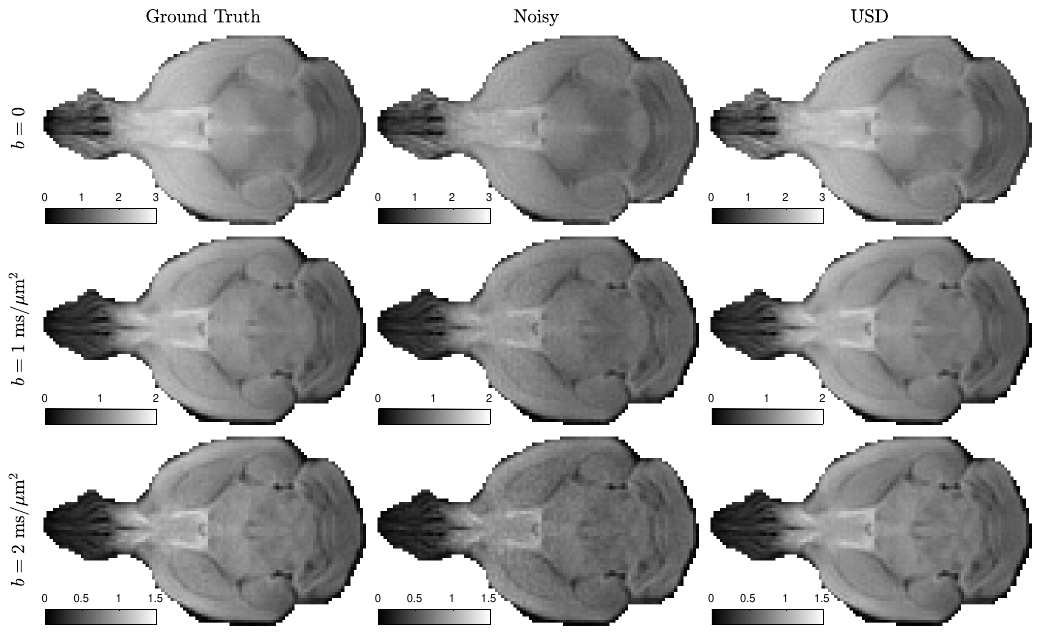}
	\caption{Demonstration of the USD denoising pipeline on dMRI data in ex vivo mouse brain (4$b=0$ images + 60 DWIs). The ground truth data was reconstructed from 402 spokes per DWI, averaged over 2 repeated measurements. The noisy data was reconstructed from 201 radial spokes per DWI, without averaging over repeats. The noise in DWIs was reduced by the USD.}
	\label{fig:mouse-dwi}
\end{figure}

\begin{figure}[t!]
\centering
	\includegraphics[width=0.99\textwidth]{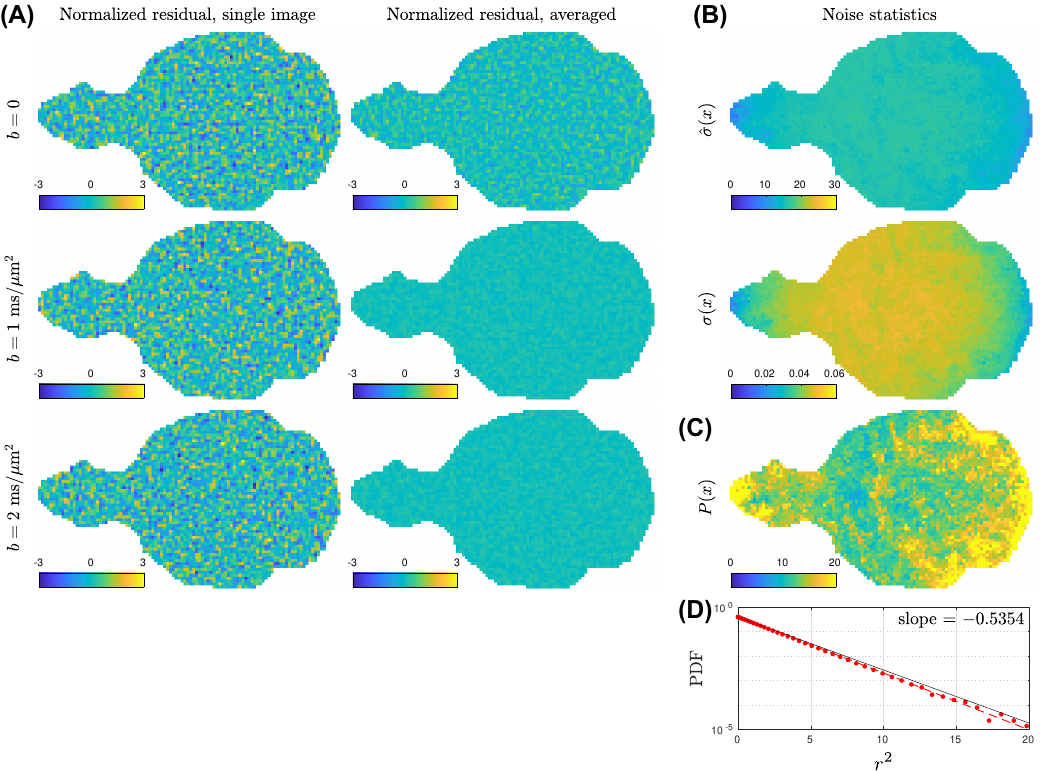}
	\caption{The noise statistics of USD in DWIs of ex vivo mouse brain (\figref{fig:mouse-dwi}). (A) The normalized residual maps $r$ in \eqref{eq:residual} had no anatomical structures, either before (left) or after (right) averaging over multiple residual maps in different gradient directions of each $b$-shell. (B) The noise maps before and after re-normalization, $\hat{\sigma}(x)$ and $\sigma(x)$, were both smooth. (C) The number $P(x)$ of signal components in PCA domain was low at the central region, where the SNR improvement $\sim\sqrt{M/P}$ was the highest. (D) The normalized image residuals $r$ were roughly normally distributed and below the reference line of slope $-1/2$ in semi-log scale, indicating that USD only removed noise.}
	\label{fig:mouse-residual}
\end{figure}

In the ex vivo mouse brain data, the USD pipeline substantially reduced the noise (\figref{fig:mouse-dwi}), especially in DWIs. 
The normalized residual maps had no anatomical structures in residuals of single images, nor in residuals averaged over multiple DWIs of each $b$-shell (\figref{fig:mouse-residual}A). The noise maps before and after re-normalization were both smooth (\figref{fig:mouse-residual}B), and the number of signal components in PCA domain was low at the central region (\figref{fig:mouse-residual}C). The histogram of normalized image residuals showed that the noise removed by USD was normally distributed up to 4 standard deviations (\figref{fig:mouse-residual}D), and its curve in the semi-log scale was below the reference line of slope $-1/2$, 
indicating that USD only removes the noise without corrupting signals. This was also supported by the absence of anatomical structure in the residual map.

\begin{figure}[t!]
\centering
	\includegraphics[width=0.99\textwidth]{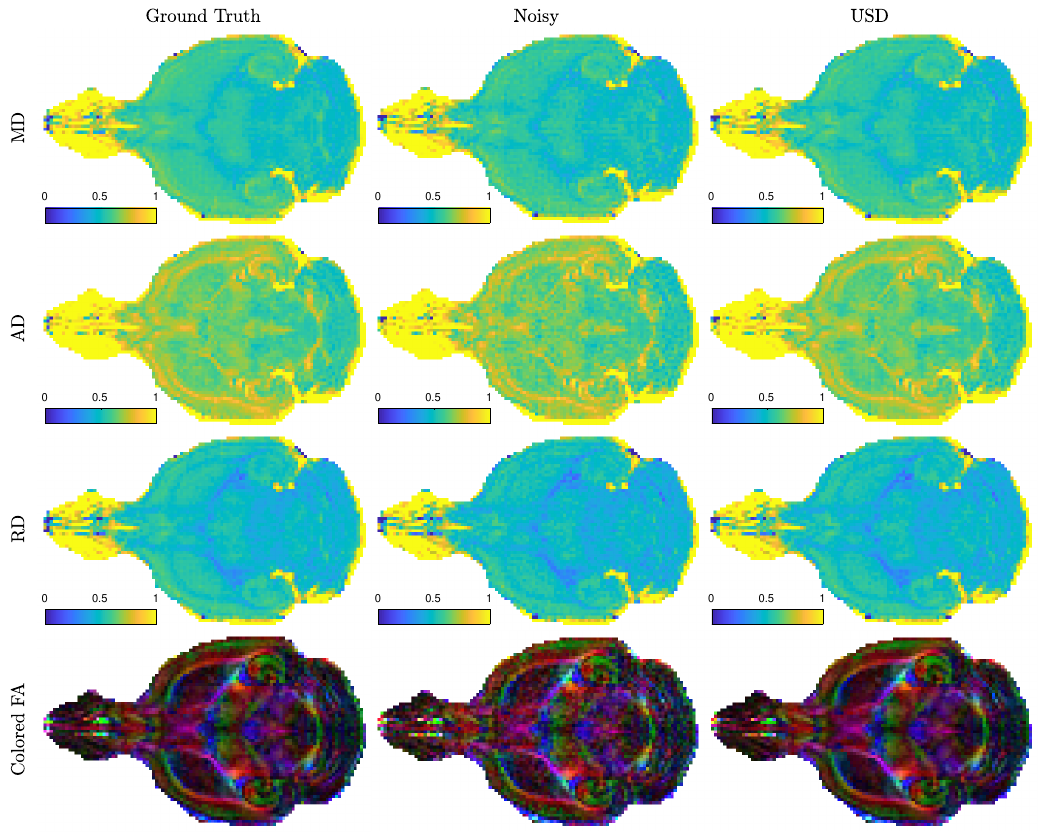}
	\caption{The effect of the USD denoising pipeline on diffusion tensor metrics (MD/AD/RD/colored FA) in ex vivo mouse brain (\figref{fig:mouse-dwi}). After denoising by using USD, the noise in colored FA map was largely reduced. Here, the colored FA map was multiplied by a factor of 2 to enhance the contrast. The units of MD, AD, and RD were $\mu$m\textsuperscript{2}/ms.}
	\label{fig:mouse-diffusivity}
\end{figure}

\begin{figure}[t!]
\centering
	\includegraphics[width=0.99\textwidth]{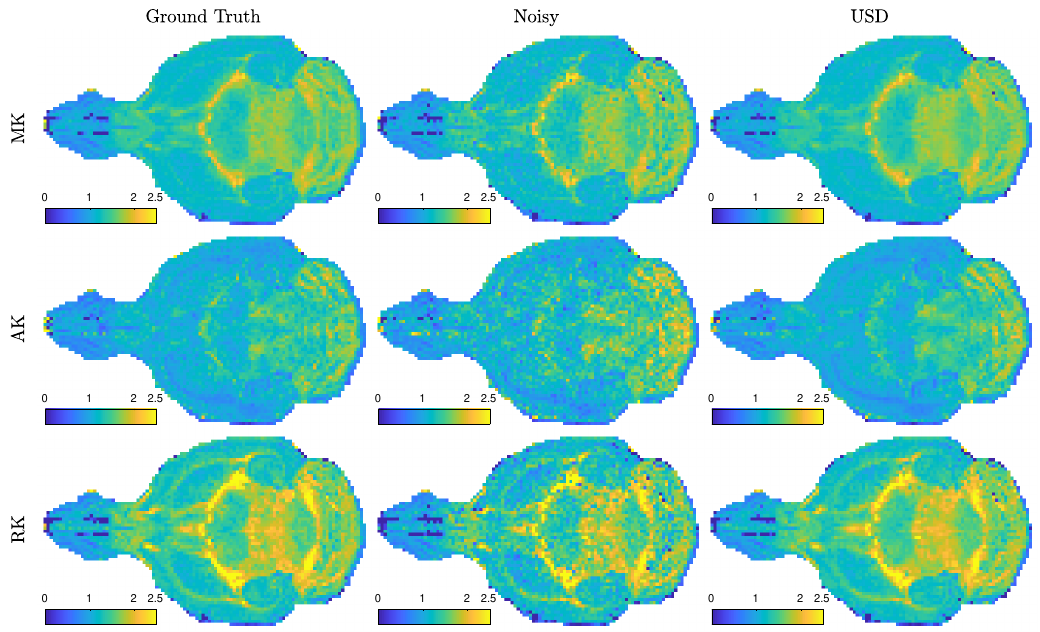}
	\caption{The effect of the USD denoising pipeline on kurtosis tensor metrics (MK/AK/RK) in ex vivo mouse brain (\figref{fig:mouse-dwi}). The kurtosis tensor was fitted by using weighted linear least square without any constraints. By applying the USD pipeline to the noisy data, the noise in MK, AK, and RK maps was largely reduced.}
	\label{fig:mouse-kurtosis}
\end{figure}

\begin{figure}[t!]
\centering
	\includegraphics[width=0.99\textwidth]{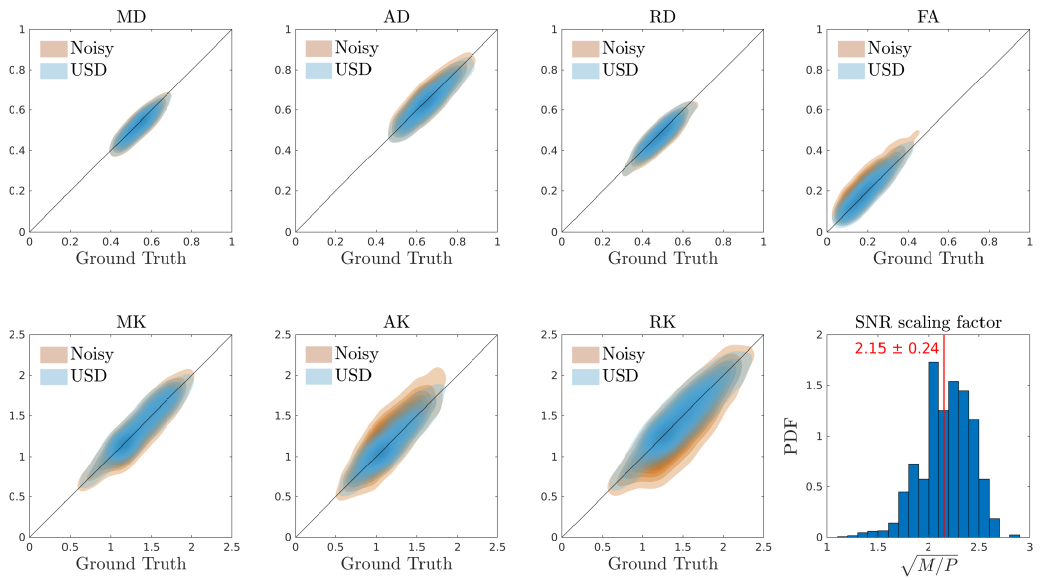}
	\caption{The effect of the USD denoising pipeline on diffusion and kurtosis tensor metrics (MD/AD/RD/FA and MK/AK/RK) and its SNR gain/scaling in an ex vivo mouse brain. In noisy data, the AD, FA, and AK were overestimated, and the RD and RK were underestimated. Denoising using USD corrected the bias, with the SNR gain $\approx2$. The units of MD, AD, and RD were $\mu$m\textsuperscript{2}/ms.}
	\label{fig:mouse-hitogram}
\end{figure}

Furthermore, USD improved the precision in parametric maps of diffusion (\figref{fig:mouse-diffusivity}) and kurtosis tensors (\figref{fig:mouse-kurtosis}), such as colored FA, MK, AK, and RK maps. The slight decrease in structural details, compared with the ground truth (402 spokes per image), was potentially due to the sub-sampling in the noisy and USD-denoised data (201 spokes per image). In particular, the eigenvalue repulsion due to the noise fluctuation lead to the overestimated AD, FA, and AK and underestimated RD and RK in the noisy mouse brain data (\figref{fig:mouse-hitogram}). This effect has been demonstrated in simulations in Ref.~\citep{koay2006rician} and Supporting Information (Figures~S1 and S2), where the diffusion signal in white matter was simulated based on the standard model \citep{jelescu2015wmtinoddi,novikov2018rotinv}. These biases in diffusion and kurtosis metrics were corrected by the USD pipeline, and the SNR was increased by a factor of $\sqrt{M/P}\approx2$ after denoising (\figref{fig:mouse-hitogram}).

\subsection{In vivo human data}
\begin{figure}[t!]
\centering
	\includegraphics[width=0.75\textwidth]{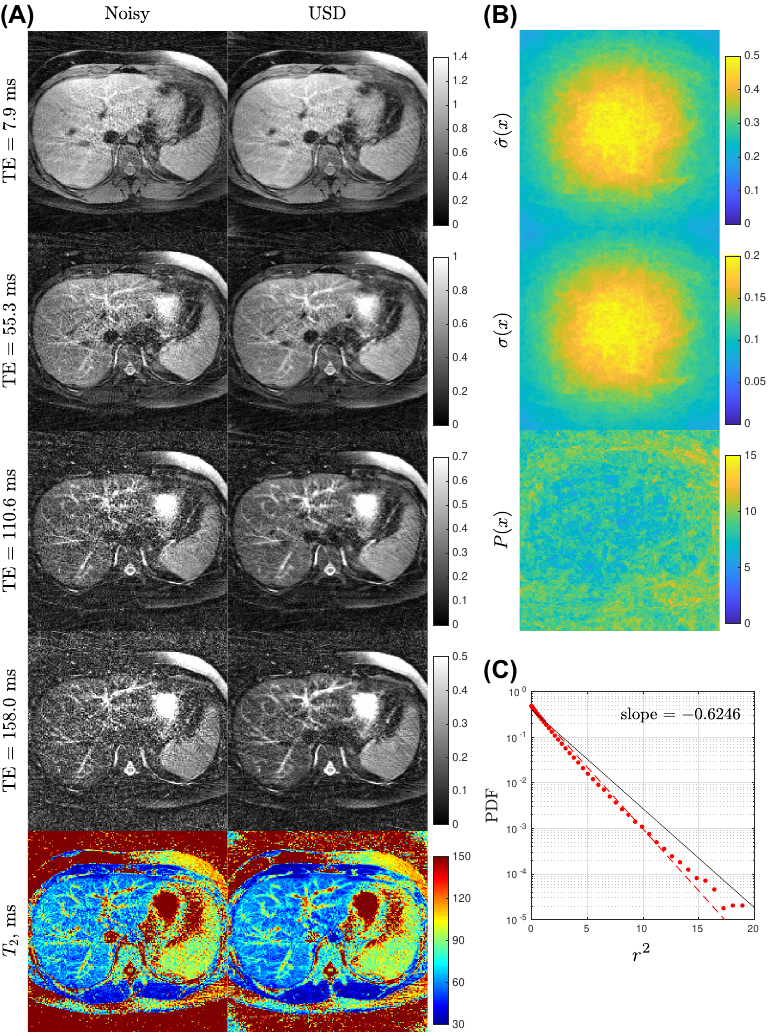}
	\caption{Demonstration of the USD denoising pipeline on T2 MRI in an in vivo human abdomen (20 T2w images). (A) The noise in T2w images and the fitted T2 map was reduced by USD. (B) The noise maps before and after re-normalization, $\hat\sigma(x)$ and $\sigma(x)$, were both smooth. And the number $P(x)$ of signal components in PCA domain was low at central region. (C) The normalized image reisdual $r$ in \eqref{eq:residual} was roughly normally distributed and below the reference line of slope -1/2 in semi-log scale, indicating that the USD only removed the noise.}
	\label{fig:human-T2}
\end{figure}

In human abdomen, the USD pipeline substantially reduced the noise in T2w images (\figref{fig:human-T2}), especially in those of TE $>$ 50 ms. The noise maps were smooth, and the number of signal components was low at the central region. The histogram of normalized image residuals in the semi-log scale was below the line of slope $-1/2$, showing that USD only removed noise without corrupting signals. In particular, the $T_2$ value in liver was 67$\pm$15 ms in noisy data and 64$\pm$12 ms in denoised data. As a reference, the liver $T_2$ value at 0.55T  was 66$\pm$6 ms in previous human studies \citep{campbell2019opportunities}.

\section{Discussion}


In this study, we propose a universal denoising pipeline applicable to any k-space sampling trajectories and demonstrate that the noise with correlation between voxels due to NUFFT for non-Cartesian MRI is removed in a numerical phantom, dMRI data of an ex vivo mouse brain, and T2 relaxation data of an in vivo human abdomen. The noise in images and parametric maps is largely reduced, the noise maps and residual maps have no anatomical structures, and the residual histogram is roughly i.i.d. Gaussian, all indicating that the USD pipeline only removes the noise without corrupting signals.

Currently we only implement USD for linear reconstruction of non-Cartesian MRI, such as the NUFFT in \eqref{eq:Ik-C-NC}. However, the nonlinear reconstruction complicates the noise statistics with intractable noise correlation. Alternatively, for nonlinear transformation, the USD can be adapted into a two-step approach: First, non-Cartesian k-space is reconstructed and denoised by using linear transformation and USD. Second, the k-space data of denoised images is re-sampled into the original non-Cartesian k-space trajectory, and further reconstructed by using a nonlinear transformation. This two-step approach potentially extends the applicability of USD to many other pipelines, such as Cartesian or non-Cartesian MRI that is highly accelerated in spatial \citep{setsompop2012blipped,dong2019tilted} and temporal domains \citep{wang2019epti}, where the challenge is that the noise correlation in coil and temporal dimensions will further increase the size of the noise covariance matrix $\Psi$ substantially. To reduce the computational load, it is possible to denoise such data by only accounting for the noise variation in all dimensions without considering the noise correlation, i.e., setting off-diagonal elements in $\Psi$ as zero. Moreover, the noise mapping given by USD could also benefit deep-learning-based imaging reconstruction and processing by providing an reliable estimate of regularization factors.

In addition to MRI, USD is potentially applicable to imaging modalities sampled in the projection space, such as PET, CT, and SPECT. With the proper treatment of Poisson-Gaussian noise statistics, it is possible to generalize the USD as a universal denoising algorithm for many other medical imaging techniques.

\section{Conclusion and outlook}
The USD pipeline successfully estimates the noise level and reduces the noise in non-Cartesian acquired data in a numerical phantom, dMRI data of an ex vivo mouse brain, and T2 relaxation data of in vivo human abdomen. Though tested only in 2d radially sampled MRI, the USD pipeline may also apply to noise removal of MRI, CT, and PET data acquired in any 2d/3d k-space/projection-space sampling scheme, as long as sufficient data redundancy is presented. The USD pipeline can be either applied before the image reconstruction or incorporated as part of it, facilitating the data under-sampling and fast imaging in future study.

\section*{Acknowledgements}\label{acknowledgements}

We would like to thank Hersh Chandarana, and Tobias Block for the discussion of T2 relaxation time mapping.
Research was supported by the Office of the Director and the National Institute of Dental and Craniofacial Research of the NIH under the award number DP5 OD031854, by the National Institute of Neurological Disorders and Stroke of the NIH under award R01 NS088040, by the National Institute of Biomedical Imaging and Bioengineering of the NIH under award number R01 EB027075, by the Office of the Director of the NIH under award number DP5 OD031854, and was performed at the Center of Advanced Imaging Innovation and Research (CAI2R, www.cai2r.net), an NIBIB Biomedical Technology Resource Center (NIH P41 EB017183).

\section*{Conflict of interest}
A patent application was submitted based on the content of the study.

\section*{Data availability statement}
The source code of the Universal Sampling Denoising pipeline will be released on our Github page (\url{https://github.com/NYU-DiffusionMRI}).

\appendix
\section{Noise level transformation}
\label{sec:noise-level-transform}
Given that the noise at a voxel $x$ in an image 
$\tilde{I}(x)$ is i.i.d. Gaussian and has a standard deviation $\hat{\sigma}(x)$, its linear transformation by an arbitrary matrix $E$,
\begin{equation}
    I(x) = \sum_{x'} E_{xx'} \, \tilde{I}(x')\,,
    \quad
    \varepsilon_x = \sum_{x'} E_{xx'} \, \tilde\varepsilon_{x'} \,, 
\end{equation}
leads to the noise correlator 
\begin{equation} \label{eq:EE}
    \langle \varepsilon_x\varepsilon^*_y\rangle = \sum_{x',y'} E_{xx'}E^*_{yy'} \langle \tilde\varepsilon_{x'}\tilde\varepsilon^*_{y'}\rangle =
    \sum_{x'} E_{xx'} E^*_{yx'} \sigma^2_0
    \equiv \sigma_0^2 (EE^H)_{xy} \,,
\end{equation}
where we used the i.i.d. property $\langle \tilde\varepsilon_{x'}\tilde\varepsilon_{y'}\rangle = \delta_{x'y'} \sigma^2_0$. 
In practice, the noise variance $\sigma_0^2$ can be estimated by the noise map $\hat{\sigma}^2(x)$ yielded by the denoising algorithm or its average $\langle\hat{\sigma}^2(x)\rangle$ over space.
Then the image $I(x)$ after applying linear transformation
has the noise of variance $\sigma^2(x)=\langle \varepsilon_x\varepsilon_x^*\rangle$, given by
\begin{equation} \label{eq:noise-linear}
    \sigma(x) = \sqrt{\left|E\cdot E^H\right|_{x,x}}\cdot\sigma_0,
\end{equation}
based on \eqref{eq:EE}. Similar conclusion has been made in Eqs.~[3] and [7] in Ref.~\citep{breuer2009gfactor}.

Substituting $E=F^H\cdot\Psi^{1/2}\cdot F$ and Fourier transform $F$ into \eqref{eq:noise-linear}, and incorporating the de-apodization scaling $c(x)$ in the last step of NUFFT, we obtain the noise level transformation in USD for each coil image in \eqref{eq:noise-usd}. Similarly, substituting $E=p^T\cdot F^H \cdot\Psi^{1/2}\cdot F$ into \eqref{eq:noise-linear}, we obtain \eqref{eq:noise-usd-comb} for the combined image.

\newpage
\section*{References}
\bibliography{references.bib}
\bibliographystyle{elsarticle-harv}
\restoregeometry

\clearpage
\setcounter{page}{1}
\setcounter{section}{0}
\section{Supporting Information}
\renewcommand{\thefigure}{S\arabic{figure}}
\setcounter{figure}{0}

\begin{figure}[h!]
\centering
	\includegraphics[width=0.8\textwidth]{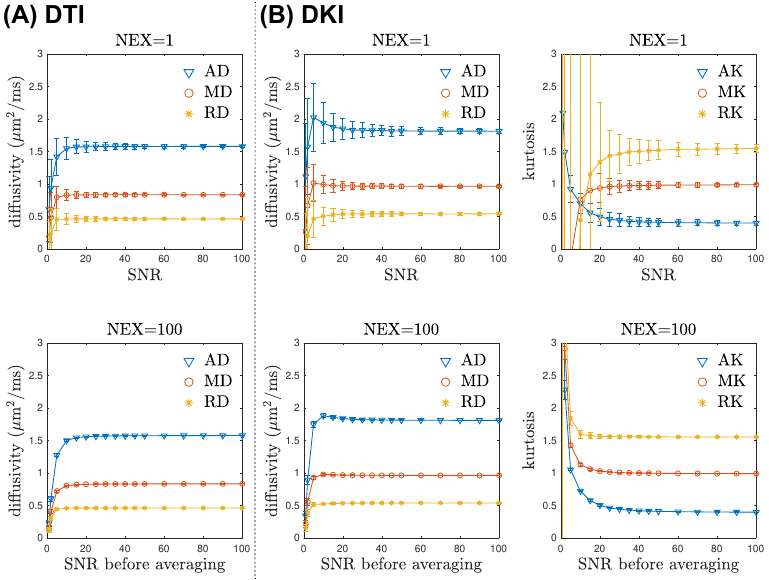}
	\caption{The effect of Rician noise on the eigenvalue repulsion and diffusion metrics in white matter. Diffusion signals 
	$
	    S = \int \text{d}\xi\,{\cal P}(\xi)\, \left[f\cdot e^{-bD_a\xi^2} + 
	    (1-f-f_\text{CSF})\cdot e^{-bD_e^\perp-b(D_e^\parallel-D_e^\perp)\xi^2} \right] +
	    f_\text{CSF}\cdot e^{-bD_\text{CSF}}
	$
	were simulated based on the Standard Model of diffusion in white matter\citep{novikov2019review} involving three compartments: intra-axonal space (stick), extra-axonal space (anisotropic Gaussian), and CSF/free water component (isotropic Gaussian),
	with parameters $D_a=2.5$ $\mu$m\textsuperscript{2}/ms, $D_e^\parallel=2$ $\mu$m\textsuperscript{2}/ms, $D_e^\perp=0.5$ $\mu$m\textsuperscript{2}/ms, $D_\text{CSF}=3$ $\mu$m\textsuperscript{2}/ms, $f=50$\%, and $f_\text{CSF}=5$\%. The fiber orientation distribution $\cal P$ was sampled in 200 directions, based on a Watson distribution of concentration parameter = 5, corresponding to a spherical harmonic coefficient $p_2=0.614$ and a dispersion angle $\theta=30.5^\circ$. The noisy data $S_n$ with Rician noise was simulated by adding the Gaussian noise ($\epsilon_r$ and $\epsilon_i$ for real and imaginary parts) to the noise-free signal $S$ via $S_n^2 = (S+\epsilon_r)^2 + \epsilon_i^2$. The signal-to-noise ratio (SNR) was defined as $S(b=0)/\sigma_g$, where $\sigma_g^2=\langle\epsilon_r^2\rangle=\langle\epsilon_i^2\rangle$. To  further explore the effect of Rician noise floor (the lower row), the magnitude of noisy data was averaged over 100 noise realizations (NEX = 100) to reduce the noise fluctuation, while the Rician noise floor remained. The simulation was repeated 1000 times; the data point and error bar were the median and the standard deviation of the repeats. 
	(A) Diffusion tensor imaging (DTI) was fitted to the data of 2 b=0 signals and DW signals of $b=1$ ms/$\mu$m\textsuperscript{2} in 30 directions by using weighted linear least square (WLLS). The eigenvalue repulsion in diffusivity was not observed (upper figure). In contrast, the Rician noise floor led to the diffusivity decrease at SNR $<$ 10 (lower figure). 
	(B) Diffusional kurtosis imaging (DKI) was fitted to the data of 2 b=0 signals and DW signals of $b=[1,2]$ ms/$\mu$m\textsuperscript{2} in [30, 60] directions by using WLLS. The eigenvalue repulsion in AD and RD was observed at SNR $<$ 20 (upper left), and the Rician noise floor led to the diffusivity decrease at SNR $<$ 10 (lower left). Furthermore, the eigenvalue repulsion in AK and RK was also observed at SNR $<$ 40 (upper right), and the Rician noise floor led to the kurtosis increase at SNR $<$ 30 (lower right).}
	\label{fig:dki-repulsion-wm}
\end{figure}

\begin{figure}[h!]
\centering
	\includegraphics[width=0.8\textwidth]{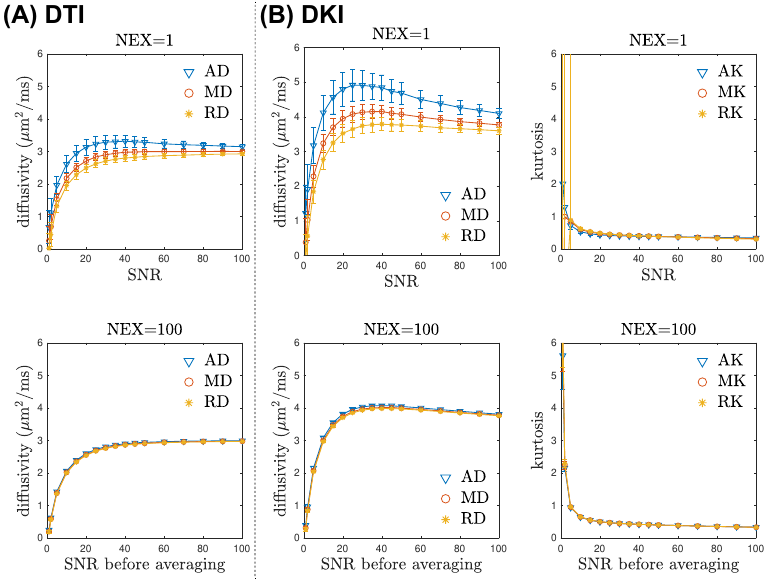}
	\caption{The effect of Rician noise on the eigenvalue repulsion and diffusion metrics in cerebrospinal fluid (CSF). Diffusion signals $S$ in CSF were simulated based on an isotropic Gaussian diffusion,
	$
	    S = e^{-bD_\text{CSF}}\,,
	$
	with the diffusivity $D_\text{CSF}=3$ $\mu$m\textsuperscript{2}/ms. The noisy data $S_n$ with Rician noise was simulated by adding the Gaussian noise ($\epsilon_r$ and $\epsilon_i$ for real and imaginary parts) to the noise-free signal $S$ via $S_n^2 = (S+\epsilon_r)^2 + \epsilon_i^2$. The signal-to-noise ratio (SNR) was defined as $S(b=0)/\sigma_g$, where $\sigma_g^2=\langle\epsilon_r^2\rangle=\langle\epsilon_i^2\rangle$. To  further explore the effect of Rician noise floor (the lower row), the magnitude of noisy data was averaged over 100 noise realizations (NEX = 100) to reduce the noise fluctuation, while the Rician noise floor remained. The simulation was repeated 1000 times; the data point and error bar were the median and the standard deviation of the repeats. 
	(A) Diffusion tensor imaging (DTI) was fitted to the data of 2 b=0 signals and DW signals of $b=1$ ms/$\mu$m\textsuperscript{2} in 30 directions by using weighted linear least square (WLLS). The eigenvalue repulsion in diffusivity was observed even at SNR = 100 (upper figure). Furthermore, the Rician noise floor led to the diffusivity decrease at SNR $<$ 40 (lower figure). 
	(B) Diffusional kurtosis imaging (DKI) was fitted to the data of 2 b=0 signals and DW signals of $b=[1,2]$ ms/$\mu$m\textsuperscript{2} in [30, 60] directions by using WLLS. The eigenvalue repulsion in AD and RD was observed even at SNR = 100 (upper left), and the Rician noise floor led to the diffusivity decrease even at SNR $<$ 30 (lower left); further, the diffusivity is overestimated at SNR $>$ 10 (cf. ground truth = $D_\text{CSF}$) potentially due to the bias of non-zero kurtosis (upper/lower right). Finally, the eigenvalue repulsion in kurtosis was not observed (upper right), and the Rician noise floor led to kurtosis increase at SNR $<$ 20 (lower right).}
	\label{fig:dki-repulsion-csf}
\end{figure}

\end{document}